\documentclass[11pt,onecolumn]{IEEEtran}
\usepackage{amsmath}
\usepackage{amssymb}
\usepackage{amsfonts}
\usepackage{epsfig}
\usepackage{subfigure}
\usepackage{calc}
\usepackage{color}
\usepackage[all]{xy}
\usepackage{bm}
\usepackage{enumerate}
\usepackage{hyperref}
\usepackage[noadjust]{cite}
\usepackage{etoolbox} 
\makeatletter
\patchcmd{\@makecaption}
{\scshape}
{}
{}
{}
\makeatletter
\patchcmd{\@makecaption}
{\\}
{.\ }
{}
{}
\makeatother

\newtheorem{defn}{Definition}
\newtheorem{thm}{Theorem}[section]
\newtheorem{cor}[thm]{Corollary}
\newtheorem{prop}{Proposition}

\newtheorem{lem}[thm]{Lemma}
\newtheorem{conj}[thm]{Conjecture}
\newtheorem{constr}[thm]{Construction}
\newtheorem{note}{Remark}
\newtheorem{example}{Example}
\newcommand{\bit}{\begin{itemize}}
\newcommand{\eit}{\end{itemize}}
\newcommand{\bcor}{\begin{cor}}
\newcommand{\ecor}{\end{cor}}
\newcommand{\beq}{\begin{equation}}
\newcommand{\eeq}{\end{equation}}
\newcommand{\beqn}{\begin{equation*}}
\newcommand{\eeqn}{\end{equation*}}
\newcommand{\bea}{\begin{eqnarray}}
\newcommand{\eea}{\end{eqnarray}}
\newcommand{\bean}{\begin{eqnarray*}}
\newcommand{\eean}{\end{eqnarray*}}
\newcommand{\ben}{\begin{enumerate}}
\newcommand{\een}{\end{enumerate}}
\newcommand{\bdefn}{\begin{defn}}
\newcommand{\edefn}{\end{defn}}
\newcommand{\bnote}{\begin{note}}
\newcommand{\enote}{\end{note}}
\newcommand{\bprop}{\begin{prop}}
\newcommand{\eprop}{\end{prop}}
\newcommand{\blem}{\begin{lem}}
\newcommand{\elem}{\end{lem}}
\newcommand{\bthm}{\begin{thm}}
\newcommand{\ethm}{\end{thm}}
\newcommand{\bconj}{\begin{conj}}
\newcommand{\econj}{\end{conj}}
\newcommand{\bconstr}{\begin{constr}}
\newcommand{\econstr}{\end{constr}}
\newcommand{\bpf}{\begin{proof}}
\newcommand{\epf}{\end{proof}}
\newcommand{\bprf}{\ \ \ {\em Proof: }}

\newcommand{\projspace}{\mbox{$ \mathbb{P}^{m-1}(\mathbb{F}_2)$}}

\begin{document}
\sloppy
\title{Binary, Shortened Projective Reed Muller Codes for Coded Private Information Retrieval} 


\author{
 \IEEEauthorblockN{Myna Vajha, Vinayak Ramkumar, and P. Vijay Kumar}\\
\IEEEauthorblockA{Department of Electrical Communication Engineering, Indian Institute of Science, Bangalore.\\
 Email: \{myna, vinram, vijay\}@ece.iisc.ernet.in} 
\thanks{Myna would like to thank the support of Visvesvaraya PhD Scheme for Electronics \& IT awarded by DEITY, Govt. of India. P. V. Kumar is also an Adjunct Research Professor at the University of Southern California.  His research is supported in part by the National Science Foundation under Grant No. 1421848 and in part by the joint UGC-ISF research program.}
}
\maketitle

\begin{abstract}
The notion of a Private Information Retrieval (PIR) code was recently introduced by Fazeli, Vardy and Yaakobi \cite{FazeliVardyYaak} who showed that this class of codes permit PIR at reduced levels of storage overhead in comparison with replicated-server PIR.  In the present paper, the construction of an $(n,k)$ $\tau$-server binary, linear PIR code having parameters $n = \sum\limits_{i = 0}^{\ell} {m \choose i}$, $k = {m \choose \ell}$ and $\tau = 2^{\ell}$ is presented. 
These codes are obtained through homogeneous-polynomial evaluation and correspond to the binary, Projective Reed Muller (PRM) code. The construction can be extended to yield PIR codes for any $\tau$ of the form $2^{\ell}$, $2^{\ell}-1$ and any value of $k$, through a combination of single-symbol puncturing and shortening of the PRM code.  Each of these code constructions above, have smaller storage overhead in comparison with other PIR codes appearing in the literature. 

For the particular case of $\tau=3,4$, we show that the codes constructed here are optimal, systematic PIR codes by providing an improved lower bound on the block length $n(k, \tau)$ of a systematic PIR code.  It follows from a result by Vardy and Yaakobi \cite{VardyYaak}, that these codes also yield optimal, systematic primitive multi-set $(n, k, \tau)_B$ batch codes for $\tau=3,4$.  The PIR code constructions presented here also yield upper bounds on the generalized Hamming weights of binary PRM codes.
\end{abstract}


\section{Introduction}
Private Information Retrieval (PIR) refers to the retrieval of data from a database without revealing information about the data being retrieved to the servers. Considering $Q_J$ as the set of queries sent to the database in order to retrieve a symbol $X_J$ whose index in the database is given by random variable $J$, we require the mutual information $I(Q_J; J)$ to be zero. The PIR problem was first introduced by Chor et al. in \cite{ChorGoldreichKushilevitzSudan_ACM} who showed that the communication complexity needs be of order $\Omega(B)$ when a single server with database of size $B$ is employed. To reduce communication complexity, the authors of \cite{ChorGoldreichKushilevitzSudan_ACM} introduced the model of non-communicating servers that store replicas of the same database and proposed algorithms for achieving PIR.  
On restricting to replicated server setting, the PIR algorithms require storage overhead to be $\ge 2$. In \cite{ShahRR14} the idea of erasure coding across PIR servers was introduced, but the metric of interest there was the amount of data downloaded and not the storage overhead. In \cite{AugotLS14}, PIR schemes based on locally-decodable codes are discussed. Coded-PIR was further explored in \cite{ChanHY15} in which the trade-off between download and storage overhead is studied.  

In \cite{FazeliVardyYaak}, \cite{FazeliVardyYaakISIT} Fazeli, Vardy and Yaakobi came up with the notion of PIR codes to achieve low storage overhead. Given an $(n,k)$ $\tau$-server PIR code, where $n$ denotes the number of servers with each server storing $\frac{B}{k}$ coded symbols, the authors provide an algorithm to achieve PIR using any existing $\tau$-replicated server protocol. An $(n,k)$ $\tau$-server PIR code, is an $(n,k)$ linear code such that for every message symbol $m_i, \ i \in [k]$, there are $\tau$ disjoint recovery sets $R_{it} \ \forall t \in [\tau]$ such that: $m_i = \sum\limits_{j \in R_{it}} c_j \ \ \forall t \in [\tau]$,
where $\underline{c} = (c_1, \cdots, c_n)$ is a codeword.
By disjoint recovery sets, it is meant that $R_{it_1} \cap R_{it_2} = \phi$ whenever $t_1 \ne t_2$ and for any $i \in [k]$. For a PIR code with $c_j = m_i$, the singleton set $\{j\}$ can itself act as a recovery set for $m_i$. Thus in the case of a systematic PIR code, every message symbol has at least one recovery set of size $1$.
\begin{figure}[h!]
	\vspace{-8pt}
	\begin{center}
		\includegraphics[width = 8.0cm]{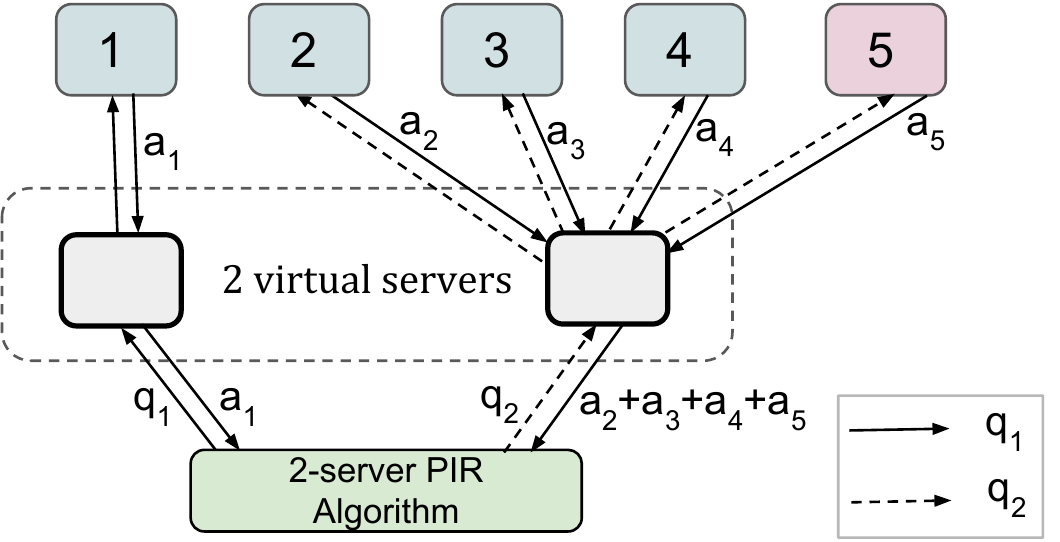}
		\caption{The working of the $(5, 4)$ 2-server PIR code described in Example~\ref{eg:2server} is illustrated here.} 
		\label{fig:2server}
	\end{center}
	\vspace{-12pt}
\end{figure}

\begin{example} \label{eg:2server}
The working of a PIR code (see \cite{FazeliVardyYaak}) is explained through an example that is illustrated in Fig.~\ref{fig:2server}.  A database of size $4B$ symbols is partitioned into the $4$ subsets $\{x_{ij} \mid j \in [B]\}_{i=1}^4$ and the $i$th subset is stored on the server numbered $i$.  The $5$th server stores $B$ symbols, each of which is the modulo-$2$ sum of the corresponding contents of the $4$ severs.   Let $Q$ and $A$ be the query and answer functions for a replicated $2$-server PIR algorithm.   In order to retrieve $x_{1j}$, queries $q_1 = Q(1,j)$, $q_2=Q(2,j)$ are generated.  The queries $q_1,q_2$ are respectively sent to the server sets corresponding to the two recovery sets $R_{11} = \{1\}$ and $R_{12} = \{2,3,4,5\}$ for message symbol $1$ in the PIR code.   Let $\{a_i\}_{i=1}^5$ be the corresponding responses, where $a_1 = A(q_1,x_1)$ and $a_i = A(q_2, x_i), \forall i \in R_{12}$.  This algorithm assumes linearity of function $A$ in its second parameter, that results in $\sum_{i \in R_{12}} A(q_2, x_i) = A(q_2, x_2+x_3+x_4+x_5) = A(q_2, x_1)$. The PIR algorithm determines $x_{1j}$ from $A(q_1, x_1)=a_1$ and $A(q_2, x_1)$. \\
\end{example}

In \cite{FazeliVardyYaak}, several PIR code constructions were proposed and connections with locally recoverable codes were made. In \cite{RaoVardy}, the authors prove a $\Omega(\sqrt{k})$ lower bound on the redundancy of an $(n,k)$ $\tau$-server PIR code and showed that this matches with the $\mathcal{O}(\sqrt{k})$ upper bound that follows from the PIR constructions in \cite{FazeliVardyYaak}.  PIR array codes are also introduced in \cite{FazeliVardyYaak}, and \cite{BlackburnEtzion}, \cite{ZhangWangWeiGe} are two recent works in that direction. In \cite{VardyYaak}, primitive multi-set batch code constructions were given using PIR codes. A $(n,k)$ linear code is called a $(n,k,\tau)_B$ primitive multi-set batch code if for any collection of $\tau$ message symbols $\underline{i} = (i_1, \cdots, i_{\tau})$ with repetition permitted, for all $t \in [\tau]$, there exists a recovery set $R_t$ for symbol $i_t$, such that $R_{t_1} \cap R_{t_2} = \phi$. In \cite{VardyYaak} it is shown that for $\tau=3,4$ optimal, systematic, PIR codes are also optimal, systematic primitive multi-set batch codes.
 
\subsection{Contributions} 
In the present paper, constructions for systematic PIR codes for $\tau$ of the form $2^{\ell}, 2^{\ell}-1$, are provided by appropriately shortening a PRM code and it is shown that these codes have lower storage overhead (smaller block lengths) in comparison with known codes\cite{FazeliVardyYaak}. A lower bound on the block length of a systematic PIR code is presented and for $\tau = 3,4$, the codes constructed here, are shown to be optimal with respect to this bound.
\subsection{Organization}
Section \ref{sec:rm_mld} presents a primer on Reed Muller (RM) codes. Binary PRM codes are introduced in Section \ref{sec:SPRM} and it is shown that this class yields efficient PIR codes.  In Section \ref{sec:subset}, a support set viewpoint of PRM codes is presented and used in Section \ref{sec:any_k}, to provide constructions of PIR codes for any $k$.  Upper bounds on the generalized Hamming weights of binary PRM codes, obtained as a by-product, appear in Section \ref{sec:any_k}. In Section \ref{sec:bounds}, an improved lower bound for systematic PIR codes is presented and used in Section~\ref{sec:optimality}, to prove optimality of the constructions for $\tau=3,4$. 

We use $[a,b]$ to denote $\{a,a+1, \cdots ,b-1,b\}$, $[a]=[1,a]$, $(a,b] = [a,b] \setminus \{a\}$ and $[a,b) = [a,b] \setminus \{b\}$. 

\section{Reed Muller Code}\label{sec:rm_mld}

A codeword in a RM code $RM(r,m)$ \cite{Muller} is a vector of $2^m$ evaluations of a polynomial 
\bea
f(x_1, \cdots, x_m) = \sum\limits_{ S \subseteq [m], |S| \le r} a_{S} \prod\limits_{i \in S} x_i, \ \ \ a_{S} \in \mathbb{F}_2, \label{eq:RM} 
\eea
of degree $\le r$ over $\mathbb{F}_2$, in the $m$ binary variables $x_i$. The coefficients $a_{S}$ represent the information symbols. The $RM(r,m)$ code has parameters: $n = 2^m$ and $k = \sum\limits_{i=0}^r {m \choose i}$.

A sequential decoding algorithm to recover message symbols is provided in \cite{Reed}.  The coefficients corresponding to the highest-degree monomials are decoded first according to: 
\bea
\label{eq:mld}
a_R &=& \sum\limits_{x_R \in \mathbb{F}_2^r} f(x_R, \underline{b}) \ \text{for any } R \subseteq [m] \text{ and } |R|=r, \
\eea
where $x_R \in \mathbb{F}_2^r$, refers to the collection of variables $(x_i | \ \forall i \in R)$ and $\underline{b} \in \mathbb{F}_2^{m-r}$ corresponds to a particular value of $x_{[m] \setminus R}$. There are $2^{m-r}$ possible values $\underline{b}$ can take resulting in 
$2^{m-r}$ recovery equations. On considering recovery equations corresponding to  $\underline{b}_1$ and $\underline{b}_2$ where $\underline{b}_1 \ne \underline{b}_2$ for a given message symbol $a_R$, it can be seen that the indices of code symbols involved are disjoint.
Therefore any $a_R$  $ \text{for all } R \subseteq [m] \text{ and } |R|=r $, can be recovered from $2^{m-r}$ disjoint recovery sets. 
Having recovered the coefficient of the highest-degree monomial terms, the contribution of these highest-degree terms is then subtracted out, leaving us with a Boolean function of lesser degree and the process is then repeated with this lesser degree.  

\section{The Projective Reed Muller Code Construction}\label{sec:SPRM}
On account of the sequential nature of the recovery algorithm, more information is needed during the recovery of lower-degree coefficients in comparison with the coefficients of the degree-$r$ terms.  To gain access to a message symbol corresponding to a degree $i < r$ term, all the message symbols corresponding to degree $>i$ have to be previously determined. 

Clearly, this can be avoided if the polynomials appearing in \eqref{eq:RM}, were restricted to be homogeneous, i.e., the coefficients of all the lower-degree monomial terms are set equal to zero.  The restriction of evaluation to homogeneous polynomials takes us from the setting of conventional and affine RM codes to the setting of Projective Reed-Muller (PRM) codes. \\

Projective Reed-Muller (PRM) codes over the field $\mathbb{F}_q$ were introduced in \cite{Lachaud86}. A codeword in the $\text{PRM}(r, m-1)$ code corresponds to evaluations of a homogeneous polynomial of degree $r$ at a specifically-chosen representative of each of the points in the projective space $\mathbb{P}^{m-1}(\mathbb{F}_q)$.  We note however, that in the projective space \projspace, each point in projective space has just a single unique representative with $m$ components. While the block length of a binary $\text{PRM}(r, m-1)$ code is nominally equal to $2^m-1$, the evaluation of a homogeneous polynomial of degree $r$ at any coordinate $\underline{x}$ with $\text{supp}(\underline{x}) < r$ gives the value $0$. Hence, these coordinates can be deleted from the binary $\text{PRM}(r, m-1)$ code to obtain a shortened version. From now on when we refer to $\text{PRM}(r,m-1)$ code, its the shortened binary version that we refer to. It follows that the code $\text{PRM}(r,m-1)$ has block length $n = \sum\limits_{i = r}^m {m \choose i}$ and dimension $k = {m \choose r} $. \\

%
Each message symbol in the PRM code can be recovered by the same method used to recover degree-$r$ terms in the Reed  Muller code as shown in \eqref{eq:mld}.  In the recovery equation for message symbol $a_R$ given by the vector $\underline{b}$, it can be verified that there is at least one element in the summation in \eqref{eq:mld}.  This ensures that there are $\tau = 2^{m-r}$ disjoint recovery sets for the retrieval of any message symbol.  Hence the $\text{PRM}(r,m-1)$ code is a $(n,k)$, $\tau$-server PIR code, where
\bean
n = \sum\limits_{i = r}^m {m \choose i} , \  k = {m \choose r} \text{ and } 
 \tau = 2^{m-r}.
 \eean
Additionally, the recovery equation corresponding to $\underline{b} = \underline{0}$ for any message symbol $a_R$, gives us $a_R = f(\underline{1}_R)$, where $\underline{1}_R$ is a binary vector with support set $R$. This establishes that the code $\text{PRM}(r,m-1)$ is a systematic code.  \\

\begin{example}
Consider the code $\text{PRM}(2,3)$. This code has parameters $(n = 11, \ k = 6, \ \tau = 4)$. A code vector in $\text{PRM}(2,3)$ corresponds to the evaluation of polynomials of form $f(\underline{x}) = a_{12} x_1x_2 + a_{13} x_1 x_3 + a_{14} x_1 x_4 +  a_{23} x_2 x_3 + a_{24} x_2 x_4 + a_{34} x_3 x_4$  of degree $2$ in $4$ variables at points $\underline{x} = (x_1, x_2, x_3, x_4)$ such that $w_H(\underline{x}) \ge 2$.
Next, consider the recovery of the coefficient $a_{12}$, of $x_1x_2$.  This coefficient can be recovered by fixing $(b_3, b_4)$ and summing over the support of the corresponding recovery sets as shown below. There are $4$ possible values of $(b_3, b_4)$ and hence $4$ disjoint recovery sets for $a_{12}$.
\bean
a_{12} &=& \sum\limits_{x_1, x_2} f(x_1x_2b_3b_4)\\
&=& f(1100) \\
&=& f(0110) + f(1010) + f(1110) \\ 
&=& f(0101) + f(1001) + f(1101) \\
&=& f(0011) + f(0111) + f(1011) + f(1111).
\eean
Generator matrix (permuted) for the $PRM(2,3)$ code is given by:

\bean
G &=& \left[
\begin{array}{ccccccccccc}
	1 & 0 & 0 & 0 & 0 & 0 & 1 & 1 & 0 & 0 & 1\\
	0 & 1 & 0 & 0 & 0 & 0 & 0 & 1 & 1 & 0 & 1\\
	0 & 0 & 1 & 0 & 0 & 0 & 1 & 0 & 1 & 0 & 1\\
	0 & 0 & 0 & 1 & 0 & 0 & 0 & 1 & 0 & 1 & 1\\
	0 & 0 & 0 & 0 & 1 & 0 & 1 & 0 & 0 & 1 & 1\\
	0 & 0 & 0 & 0 & 0 & 1 & 0 & 0 & 1 & 1 & 1\\
\end{array}
\right]
\eean
\end{example}
The $\text{PRM}$ codes have in general, non-uniform information-symbol locality. For instance, in the example above, there are $2$  sets with locality $3$ and $1$ set with locality $4$. Each recovery set $R_{\underline{b}}$ is naturally associated to a specific vector $\underline{b} \in \mathbb{F}_2^{m-r}$.
Let $w_b$ denote the Hamming weight of the vector $\underline{b}$.    There are ${m-r \choose w_b}$ recovery sets with cardinality $R_{\underline{b}}$ and 
\bean
\left| R_{\underline{b}} \right| & = & \left\{ \begin{array}{rl} \sum\limits_{i=0}^{w_b} {r \choose r-w_b+i} &  w_b < r, \\
2^r &  w_b \ge r. \end{array} \right.
\eean
%
Since $| \cup_{\underline{b} \in \mathbb{F}_2^r} R_{\underline{b}}| = n$, it follows that all code symbols participate in the $\tau$ recovery equations corresponding to each of the message symbols.


\section{Support-Set Viewpoint of PRM Codes}\label{sec:subset}


Each code symbol of an $\text{PRM}(r,m-1)$ code, is indexed by a vector $\underline{x} \in \mathbb{F}_2^m$ with $w_H(\underline{x}) \ge r$.  Since each of these vectors is uniquely represented by its support set, each code symbol can equivalently, be indexed by a subset of $[m]$ of size $\ge r$.  
Our aim in the next section, is to construct PIR codes for other values of $k$.  Our approach is to consider shortened versions of the PRM code, obtained by judiciously setting certain of the message symbols to zero.  When we set a certain message symbol to equal zero, the corresponding code symbol (since the code is systematic) is automatically set equal to zero.  But if a set of message coefficients is set equal to zero, it turns out that certain other code symbols are forced to be equal to zero as well.   This results in a shortened code having smaller block length.  The shortened codes are also PIR codes for exactly the same reason as is the parent PRM code. The shorter block length makes these codes more efficient as can be seen from the table \ref{table:sprm_params} of the parameters of the PIR codes so constructed.   We explain this last point in greater detail below.  \\

For $S$ a subset of $[m]$, we will for the sake of brevity, write $f(S)$ in pace of $f(\underline{1}_S)$.  For example, when $m=5$, we will write $f(\{1,2,5\})$ in place of $f(11001)$.
Next, let $R_i, \forall i \in \left[ {m \choose r} \right] $ represent the ${m \choose r}$, $r$-element subsets of $[m]$.  We note that for any subset $S \subseteq [m]$, we have that   
\bean f(S) = \sum\limits_{\forall R_i \subseteq S} f(R_i). \eean
For example, $\text{PRM}(2,3)$ code has $f(\{1,2,4\}) = f(\{1,2\}) + f(\{1,4\}) + f(\{2,4\})$.\\

It follows that if we set $f(R_i) = 0$, by setting the corresponding message coefficients to be equal to zero, $\forall R_i \subseteq S$, then $f(S) = 0$.  Thus if we shorten the PRM code by setting all message coefficients corresponding to $r$-element subsets of a fixed set $S$ to zero, then the shortening process will result in the deletion of the coordinate corresponding to the support set $S$ as well. 

\section{Constructions for any $k$ and $\tau = 2^{\ell}, 2^{\ell}-1$}\label{sec:any_k}
In this section we provide constructions for $\tau$ of the form $2^{\ell}$ for any $k$.  Each of these codes will also turn out to be systematic.  It is straightforward to show (see \cite{FazeliVardyYaak})  that if a systematic $(n,k)$, $\tau$-server PIR code is punctured by deleting a parity-check symbol, one will obtain a systematic $(n-1,k)$, $(\tau-1)$-server PIR code.  Thus our constructions for $(n,k)$, $2^{\ell}$-server PIR codes, can be punctured to yield constructions for $\tau=2^{\ell}-1$ as well. 

In this section, we will show how one can make use of the support-set viewpoint of a PRM code to shorten the code to obtain PIR codes for values of $k$ other than of the form ${m \choose \ell}$.  
To construct a PIR code for $k \in ({m-1 \choose \ell}, {m \choose \ell})$ and $\tau = 2^{\ell}$, we consider a $\text{PRM}(r, m-1)$ code, where $r = m-\ell$ and set $\gamma = {m \choose \ell} - k$, message symbols to zero to obtain the shortened Projective Reed Muller code $\text{SPRM}(r, m-1, \gamma)$ for $0 \le \gamma \le {m-1 \choose \ell-1}$. Considering $\gamma'$ as the reduction in block length on shortening $\text{PRM}(r, m-1)$ by $\gamma$, we get $n = \sum\limits_{i \in [0, \ell]} {m \choose i} - \gamma'.$It is clear that $\gamma' \ge \gamma$. 

We first show in Lemma:\ref{lem:tau_retain} that irrespective of setting any of the $\gamma$ message symbols to zero, $\tau$ is still retained. We then give an algorithm to judiciously pick the $\gamma$ message symbols to get a block length reduction of $\gamma'$ in Theorem \ref{thm:shorten3}. \\

\begin{lem}
	\label{lem:tau_retain}
	On shortening a $PRM(r,m-1)$ code by setting any $\gamma$ message symbols to zero, the resultant code retains $\tau = 2^{m-r}$ disjoint recovery sets.
\end{lem}
\bprf Consider $f(R_j), \forall j \in [\gamma]$ as the $\gamma$ message symbols that are set to zero. Any recovery equation for a left out symbol $f(R_i)$ for $i \in [\gamma+1, {m \choose r}]$ given below has $f(R_i \cup S)$ as an element.
\bean
\label{eq:mld_set}
f(R_i) = \sum\limits_{R_0 \subseteq R_i} f(R_0 \cup S) \ \ \ \forall S \subseteq [m] \setminus R_i.
\eean
 
It is clear to see that $f(R_i \cup S)$ cannot be deleted when $f(R_i)$ is not set to 0. This shows that for any $S \in [m] \setminus R_i$ we have at-least one element in the recovery equation, resulting in $\tau = 2^{\ell}$.
\ \\
\begin{thm}
	\label{thm:shorten1}
	For $\gamma = {r + t \choose r}$ for all $t \in [0, \ell-1]$, $\gamma' = \sum\limits_{i = 0}^t {r+ t \choose r+i}$ is possible.
\end{thm}
\bprf
Consider a $r + t$ element subset $T$ of $[m]$ and shorten $\text{PRM}(r, m-1)$ by setting the $\gamma$ message symbols corresponding to all the $r$-element subsets of $T$ as zero. By doing this, we can also delete code symbols corresponding to the subsets of $T$ with cardinality $\ge r$. This gives a reduction of $\gamma' = \sum\limits_{i=0}^{t} {r+t \choose r+i}$.   
\ \\
For the case of $\text{PRM}(2,4)$ code, Theorem \ref{thm:shorten1} gives the codes $\text{SPRM}(2,4, \gamma)$ for $\gamma = 1,3, 6$ with $\gamma'$ as 1,4,11 respectively. On setting $t = \ell -1$ in Theorem \ref{thm:shorten1} we get the parameters of $\text{SPRM}\big(r, m-1, {m-1 \choose r}\big)$ code as $k = {m-1 \choose \ell}$, $n = \sum\limits_{i=0}^{\ell} {m-1 \choose i}$. These parameters are same as that of $\text{PRM}(r-1, m-2)$. Therefore, we do not restrict to $\gamma < {m-1 \choose \ell}$ in the next theorems as this shortening algorithm seamlessly goes from PRM$(r, m-1)$ to PRM$(r-1, m-2)$.\\

\begin{thm}
	\label{thm:shorten2}
	For	$\gamma = \sum\limits_{i = 0}^{\rho_{t} - 1} {r+t-i \choose r-i}$ for any $t \in [0, \ell-1]$ and $\rho_t \in [1, r]$, $\gamma' = \sum\limits_{j=0}^{t} \sum\limits_{i = 0}^{\rho_{t}-1} {r+t-i \choose r+j-i}$ is possible.
\end{thm}
\bprf
Consider the set $S = [1, r+t+1]$ and the  $(r+t)$-element subsets $S_i = S \setminus \{r+t+1-i\}$, $\forall i\in [0,r+t]$. 

Consider $\rho_{t}$ such $(r+t)$-element sets $ \mathbb{P} = \big \{ S_i, \ \forall i \in [0,\rho_{t}-1] \big \}$ where $\rho_{t} \le r$ and shorten $PRM(r,m-1)$ by setting message symbols corresponding to all the distinct $r$ element subsets of sets in $\mathbb{P}$. This gives $\gamma = \sum\limits_{i = 0}^{\rho_{t} - 1} {r+t-i \choose r-i}.$

In this case we can delete all the code symbols corresponding to subsets of sets in $\mathbb{P}$ with cardinality $\ge r$ giving a reduction of $\gamma' = \sum\limits_{j=0}^{t} \sum\limits_{i = 0}^{\rho_{t}-1} {r+t-i \choose r+j-i}$ resulting in the theorem.

\ \\
For $\rho_t = 1$, Theorem \ref{thm:shorten2} falls back to the case of Theorem \ref{thm:shorten1}. Now by picking $\rho_t = 2$ for $PRM(2,4)$ code in Theorem \ref{thm:shorten2} we get the $\text{SPRM}(2,4,\gamma)$ code for $\gamma = 2, 5, 9$ with $\gamma' = 2, 7, 18$ respectively. 
We essentially extend the same idea in the next theorem to give constructions for any $k$.\\
\begin{thm}
	\label{thm:shorten3}
	For any $\gamma  \in \big[0, {m \choose \ell}\big)$, $\gamma$ can be uniquely represented using a vector $(\rho_{\ell-1}, \cdots \rho_0)$ with $\rho_i \ge 0, \forall i \in [0, \ell-1]$ and $\sum\limits_{i=0}^{\ell-1} \rho_i \le r$ as 
    \bean 
    \gamma = \sum\limits_{t = 0}^{\ell - 1} h(\rho_t, r_t, t) \ \ \ \text{ where, } h(p, r, t) = \begin{cases}
    	\sum\limits_{i = 0}^{p-1} { r+t-i \choose r-i} & p > 0\\
    	0 & p = 0 
    \end{cases} \ \text{ and } r_t = r - \small { \sum\limits_{q > t}^{\ell-1} \rho_{q}}.
    \eean
    Then for $\text{SPRM}(r, m-1, \gamma)$, reduction of 
    \bean 
    \gamma' = \sum\limits_{t=0}^{\ell - 1} h_1(r_t, t) \ \ \ \text{ where, } h_1(r, t) = \begin{cases}
		\sum\limits_{j = 0 }^{t} \sum\limits_{i=0}^{\rho_t - 1} {r+t-i \choose r+j-i} & \rho_t > 0\\
		0 & \rho_t = 0
		\end{cases}
		\eean
		is possible.
\end{thm}
 \bprf  Lets recursively define 
\bean
\gamma_t = \begin{cases}
	\gamma & t = \ell -1, \\
	\gamma_{t+1} - h( \rho_{t+1}, r_{t+1}, \ t+1) & 0 \le t < \ell-1.
\end{cases}
\eean
We determine $\rho_t$ as shown below by the index $p \in [0, r_t]$ of the interval in which $\gamma_t$ lies.
\bean
\rho_t = p \text{ such that }  \gamma_t \in \Big[ \ h(p,r_t,t), \ h(p+1, r_t, t) \ \Big).
\eean
For $t = \ell-1$, $\gamma < h(r+1, r, \ell-1) = {r+\ell \choose r} = {m \choose r}$. One can always find an interval in which $\gamma_t$ lies, otherwise we have $\gamma_t \ge h(r_t+1, r_t, t) = {r_t + t + 1 \choose r_t}$. This gives that
\bean
\gamma_{t+1} &\ge&  h( \rho_{t+1}, r_{t+1}, t+1) + {r_t + t + 1 \choose r_t}\\
&=& h(\rho_{t+1}+1, r_{t+1}, t+1) \ \{\text{  as } r_t = r_{t+1} - \rho_{t+1} \}.
\eean
This is a contradiction on definition of $\rho_{t+1}$. So we can always find an index $p \in [0,r_t]$ for $\rho_t$. We start by defining the global set as $S_0^{\ell} = [m]$ and define $\rho_{\ell} = 0$. For the set $S_i^{j}$, $j$ is  the number of elements in the set. Now we recursively define sets,
\bea
\label{eq:shorten_sets}
S_i^{r+t-1}=S_{\rho_t}^{r+t} \setminus \{r_{t-1}+t-i \},\ \forall i \in [0,r_{t-1}+t-1]
\eea
$\forall t \in [1, \ell]$. It is clear to see that $|S_i^{j} \cap S_{i'}^{j}| = j-1$ for the sets defined by \ref{eq:shorten_sets}. By picking $\rho_t$, $(r+t)$-element sets, we get $\mathbb{P} = \Big\{S_i^{r+t}, \forall t \in [0, \ell-1], \forall i\in[0,\rho_t-1] \Big\}$. It can be seen that $S_i^{r+t} \nsubseteq S_{i'}^{r+t'} $ for all $t' > t$ and $i'  \in [0,\rho_{t'}-1]$. Here, $\rho_t$ corresponds to the number of $r+t$ element sets that are not already subsets of larger cardinality sets in $\mathbb{P}$. Now by setting all the message symbols corresponding to distinct $r$-element subsets of sets in $\mathbb{P}$ to zero we get a count of $\gamma$.
Now we can delete symbols corresponding to all subsets of sets in $\mathbb{P}$ with cardinality $\ge r$. This gives us the reduction  $\gamma'$ as stated.
\ \\
Theorem \ref{thm:shorten2} is a special case of Theorem \ref{thm:shorten3}, where $\gamma$ is represented by single weight $\underline{\rho}$ vector. This can be seen in Table~\ref{table:SPRM_2_4}.

\begin{table}[h!]
	\begin{center}	
		\bean
		\begin{array}{|c|c|c|c|c|c|}
			\hline \gamma & \underline{\rho} &  \mathbb{P} & \gamma' & k & n \\ \hline
			0 & (0,0,0) & \phi & 0 & 10 &  26 \\ \hline
			1 & (0,0,1) & \{1,2\} & 1 & 9 &  25 \\ \hline
			2 & (0,0,2) & \{1,2\}, \{1,3\} &  2 & 8 & 24 \\ \hline
			3 & (0,1,0) & \{1,2,3\} & 4 & 7 &  22 \\ \hline
			4 & (0,1,1) & \{1,2,3\}, \{1,4\} & 5 & 6 & 21 \\ \hline
			5 & (0,2,0) & \{1,2,3\}, \{1,2,4\} & 7 & 5 &  19 \\ \hline
			6 & (1,0,0) & \{1,2,3,4\} & 11 & 4 & 15  \\ \hline
			7 & (1,0,1) & \{1,2,3,4\}, \{1,5\} & 12 & 3 &   14\\ \hline
			8 & (1,1,0) & \{1,2,3,4\}, \{1,2,5\} & 14 & 2 &  12 \\ \hline
			9 & (2,0,0) & \{1,2,3,4\}, \{1,2,3,5\} & 18 & 1 & 8   \\ \hline
		\end{array}
		\eean
		\caption{Parameters list of $\text{SPRM}(2,4,\gamma)$ code for $\gamma \in [0, 9]$. On counting the 2-element subsets of sets in $\mathbb{P}$ gives $\gamma$ and counting subsets of cardinality $\ge2$ gives $\gamma'$.}
		\label{table:SPRM_2_4}
	\end{center}
\end{table} 
%

\subsection{Upper bounds on generalized Hamming weights of Binary PRM codes.}
The SPRM codes presented in section \ref{sec:any_k} also give upper bound on the generalized Hamming weights of $PRM$ codes defined as $d_i = \min | \text{supp}(D)|$ where $D$ is a $i$-dimensional sub code and $\text{supp}(D)$ is the union of support of all the vectors in $D$.
For a binary $PRM(r = m-\ell, m-1)$ code,
\bea
\label{eq:ghw}
d_{k - \gamma} \le n - \gamma' \text{ where } k = {m \choose r}, \ n = \sum\limits_{i = r}^m {m \choose i},
\eea for all $\gamma \in [0,k)$.  and $\gamma'$ is as given in Theorem:\ref{thm:shorten3} for a given $\gamma$.

\paragraph{$d_1$ of PRM codes}
For a $PRM(r=m-\ell,m-1)$ code there are $\tau = 2^{\ell}$ disjoint recovery sets. This ensures that any $e \le \tau-1$ erasures can be corrected. This gives 
\bean
d_1 = d_{\min} \ge 2^{\ell}.
\eean
Now by substituting $\gamma = k-1$ in eq:\ref{eq:ghw}, we get an upper bound on $d_1$. By the unique representation shown in Theorem:\ref{thm:shorten3}, $\gamma = k-1$ is represented by vector $(r,0, \cdots, 0)$. This gives:
\bea
\label{eq:d2}
\gamma' = h_1(r, \ell-1) = \sum\limits_{j=0}^{\ell-1} \sum_{i = 0}^{r - 1} {r+\ell-1-i \choose r + j-i } 
\eea
It can be noted that
\bean
\sum_{i = 0}^{r+j} {r+\ell-1-i \choose r + j-i } = {r + \ell \choose r + j} = {m \choose r+j}
\eean
Substituting the above equation in eq:\ref{eq:d2} we have
\bean
\gamma' &=& \sum\limits_{j=0}^{\ell-1} {m \choose r+j} - \sum\limits_{j=0}^{\ell-1} \sum_{i = r}^{r+j} {r+\ell-1-i \choose r + j-i } \\
&=& n-1 - \sum\limits_{j=0}^{\ell-1} \sum_{i = 0}^{j} {\ell-1-i \choose j-i } = n - 2^{\ell}.
\eean
This gives $d_1 = 2^{\ell}$.
\paragraph{$d_2$ of PRM codes}
$\gamma = k-2$ can be represented as $(r-1, 1, 0, \cdots, 0)$ when $m > r$ (i.e., $k > 1$). This gives
\bean
\gamma' &=& h_1(r-1, \ell-1) + h_1(1, \ell-2)\\
&=& h_1(r, \ell-1) - \sum\limits_{j=0}^{\ell-1} {\ell \choose j+1} + \sum\limits_{j = 0}^{\ell-2} {\ell-1 \choose j+1}\\
&=& n - 3(2)^{\ell-1}
\eean
Substituting this in \ref{eq:ghw} we get $d_2 \le 3(2)^{m-r-1}$.




\section{Bounds for systematic PIR Codes}\label{sec:bounds}

For a systematic PIR code, the generator matrix is of the form $[I \ | \ P]$, where I is the $k\times k$ identity matrix. In this section we prove a lower bound on block length $n(k, \tau)$ of a systematic $(n,k)$ $\tau$-server PIR code. This is an improvement over the lower bound provided in \cite{RaoVardy}. We show in Section:\ref{sec:optimality} that this bound is achieved for the case of $\tau = 3, 4$ by using $\text{PRM}(m-2, m-1)$ codes and their extensions.\\

\begin{thm}
	\label{thm:lb}
	For a $(n,k)$ $3$-server systematic PIR code,
	\bean
	n(k, 3) \ge k + \left\lceil \frac{\sqrt{8k+1}+1}{2} \right\rceil.
	\eean
\end{thm}
\bprf
We consider a $(n,k)$ 3-server systematic PIR code.
For this code, let $R_{i1}, R_{i2}, \{i\}$ be the 3-disjoint recovery sets corresponding to message symbol $i$ and let
$
\left[ \begin{array}{cccc}
	I_k & g_{k+1} & \cdots & g_n
\end{array} \right]
$
be the generator matrix $G$. Then,
\bean
e_i = \sum\limits_{j \in S_{i1}} e_j + \sum\limits_{j \in T_{i1}} g_j = \sum\limits_{j \in S_{i2}} e_j + \sum\limits_{j \in T_{i2}} g_j
\eean
where, $S_{i1} = R_{i1} \cap [k]$, $S_{i2} = R_{i2} \cap [k]$, $T_{i1} = R_{i1} \setminus S_{i1}$ and $T_{i2} = R_{i2} \setminus S_{i2}$.
Let us define
\bean
u_{i1} = \sum\limits_{j \in S_{i1}} e_j, && u_{i2} = \sum\limits_{j \in S_{i2}} e_j,\\
v_{i1} = \sum\limits_{j \in T_{i1}} g_j = e_i + u_{i1}, && 
v_{i2} = \sum\limits_{j \in T_{i2}} g_j = e_i + u_{i2}.
\eean
It is clear to see that,
\bean
e_i = (e_i + u_{i1}) \odot (e_i + u_{i2}) 
= v_{i1}\odot v_{i2} = \sum\limits_{\substack{\ell \in T_{i1} \\ m \in T_{i2}}} g_{\ell} \odot g_m,
\eean
where $\odot$ is the component wise product. Now consider set $X = \left\lbrace g_{k+1}, \cdots, g_n \right\rbrace$ and define the set 
\bean
X^2 = \{ g_i \odot g_j \ | \ g_i, g_j \in X \ \& \ i \ne j \}. \eean
This gives $e_i \in <X^2>$ as $T_{i1} \cap T_{i2} = \phi$. Therefore we have,
\bean
k = \dim(e_1, \cdots, e_k) = \dim(<X^2>) \le |X^2| \le {n-k \choose 2}.
\eean
This gives us the bound for $\tau = 3$.

\ \\
\begin{cor}\label{cor:lb_tau}
\bean n(k, \tau) \ge k + \left\lceil \frac{\sqrt{8k+1}+1}{2} \right\rceil + \tau - 3. \eean
\end{cor}
This corollary holds due to the fact that $n(k, \tau) \ge n(k, \tau-1) + 1$ since deletion of a column from the generator matrix will reduce $\tau$ by at most $1$ (by \cite{FazeliVardyYaak}). Applying this fact to the bound $n(k,3) \ge  k + \left\lceil \frac{\sqrt{8k+1}-1}{2}  \right\rceil$ we get Corollary \ref{cor:lb_tau}.

\section{Optimal Codes for $\tau \le 4$}\label{sec:optimality}
For $\tau=2$, $\text{PRM}(k-1, k-1)$ is the parity check code and it is optimal. To get a PIR code with dimension $k$ and $\tau = 4$, consider $\text{PRM}(m-2,m-1)$ code, with $m$ such that $k \in \big( {m-1 \choose 2}, {m \choose 2}\big]$ and $\gamma = {{m \choose 2} - k}$. By setting any $\gamma$ message symbols to be zero, we can delete the coordinates corresponding to them. This gives:
\bean
k = {m \choose 2}-\gamma, \ \ n = k + m + 1, \ \  \tau = 4. 
\eean
In fact $\text{SPRM}(m-2, m-1, \gamma)$ has the same parameters as above. This gives $n( k ,4)\le k + m +1 $. From the lower bound on block length in Corollary \ref{cor:lb_tau}, we have that $n ( k ,4) \ge  k + m + 1 $.\\ 

On puncturing $\text{SPRM}(m-2, m-1, \gamma)$ at a parity symbol we get a $(n,k)$ 3-server PIR code, where $n = k+m$ and $k = {m \choose 2} - \gamma$. This gives the upper bound $n(k,3) \le k + m$. From the lower bound in Theorem:\ref{thm:lb} we have $n(k,3) \ge k + m $. Therefore for any $k$ we have optimal systematic PIR codes for $\tau=3,4$.\\

In \cite{VardyYaak} it was shown that optimal systematic PIR codes for $\tau = 3,4$ give optimal systematic primitive multi-set batch codes. So $\text{SPRM}(m-2,m-1, \gamma)$ and its punctured version can be used as $(n = k + m +1, k = {m \choose 2} - \gamma, 4)_B$, $(n = k + m , k = {m \choose 2} - \gamma, 3)_B$ batch codes respectively.

\begin{table}[h!]
	\label{table:sprm_params}
	\begin{center}
		\begin{tabular}{||c||c | c || c|c||c|c||c|c||}
			\hline
			
			\multicolumn{1}{||c ||}{ \ \  k $\backslash$ $\tau$ }&
			\multicolumn{2}{|c ||}{3*} &
			\multicolumn{2}{|c ||}{4*} &
			\multicolumn{2}{|c ||}{8} &
			\multicolumn{2}{c||}{16} \\
			\hline
			& $n_1$ & $n_2$ & $n_1$ & $n_2$ & $n_1$ & $n_2$ & $n_1$ & $n_2$\\
			\hline \hline
			2 & 5 & 5 &  6 & 6 & 12 & 12 & 24 & 24 \\
			\hline
			3 & 6 & 6 & 7 & 7 & 14 & 14 & 28 & 28 \\
			\hline
			4 & 8 & 8 & 9 & 9 & 15 & 15 & 30 & 30 \\
			\hline
			5 & 9 & 10 & 10 & 11 & 19 & 19 & 31 & 31 \\
			\hline
			6 & 10 & 11 & 11 & 12 & 21 & 21 & 39 & 40 \\
			\hline
			7 & 12 & 12 & 13 & 13 & 22 & 23 & 43 & 43 \\
			\hline
			8 & 13 & 13 & 14 & 14 & 24 & 28 & 45 & 54 \\
			\hline
			9 & 14 & 14 & 15 & 15  & 25 & 30 & 46 & 60 \\
			\hline
			10 & 15 & 17 & 16 & 18 & 26 & 35 & 50 & 61 \\
			\hline
			11 & 17 & 19 & 18 & 20  & 30 & 37 & 52 & 67 \\
			\hline
			12 & 18 & 20 & 19 & 21 & 32 & 39 & 53 & 69 \\
			\hline
			13 & 19 & 21 & 20 & 22 & 33 & 41 & 55 & 71 \\
			\hline
			14 & 20 & 22 & 21 & 23 & 35 & 43 & 56 & 74 \\
			\hline
			15 & 21 & 23 & 22 & 24 & 36 & 44 & 57 & 80 \\
			\hline
			16 & 23 & 24 & 24 & 25 & 37 & 45 & 65 & 84 \\
			\hline
			17 & 24 & 27 & 25 & 28 & 39 & 46 & 69 & 86 \\
			\hline
			18 & 25 & 28 & 26 & 29 & 40 & 47 & 71 & 88 \\
			\hline
			19 & 26 & 29 & 27 & 30 & 41 & 48 & 72 & 90 \\
			\hline
			20 & 27 & 30 & 28 & 31 & 42 & 49 & 76 & 92 \\
			\hline
			21 & 28 & 31 & 29 & 32 & 46 & 50 & 78 & 94 \\
			\hline
			22 & 30 & 32 & 31 & 33 & 48 & 51 & 79 & 100 \\
			\hline
			23 & 31 & 33 & 32 & 34 & 49 & 52 & 81 & 104 \\
			\hline
			24 & 32 & 34 & 33 & 35 & 51 & 53 & 82 & 106 \\
			\hline
			25 & 33 & 35 & 34 & 36 & 52 & 54 & 83 & 108 \\
			\hline
			26 & 34 & 38 & 35 & 39 & 53 & 55 & 87 & 110 \\
			\hline
			27 & 35 & 39 & 36 & 40 & 55 & 56 & 89 & 112 \\
			\hline
			28 & 36 & 40 & 37 & 41 & 56 & 57 & 90 & 114 \\
			\hline
			29 & 38 & 41 &39 & 42 & 57 & 58 & 92 & 116 \\
			\hline
			30 & 39 & 42 & 40 & 43 & 58 & 59 & 93 & 118 \\
			\hline
			31 & 40 & 43 & 41 & 44 & 60 & 60 & 94 & 120 \\
			\hline
			32 & 41 & 44 & 42 & 45 & 61 & 61 & 96 & 122 \\
			\hline

		\end{tabular}
		\\
		\vspace{5 pt}
		\caption{Block length for various $k$, $\tau$. Here $n_1$ is the block length of the SPRM constructions and $n_2$ is the block length of the best known codes provided in \cite{FazeliVardyYaak}}
	\end{center}
\end{table}


\bibliographystyle{IEEEtran}
\bibliography{isit2017}

\begin{thebibliography}{10}
\providecommand{\url}[1]{#1}
\csname url@samestyle\endcsname
\providecommand{\newblock}{\relax}
\providecommand{\bibinfo}[2]{#2}
\providecommand{\BIBentrySTDinterwordspacing}{\spaceskip=0pt\relax}
\providecommand{\BIBentryALTinterwordstretchfactor}{4}
\providecommand{\BIBentryALTinterwordspacing}{\spaceskip=\fontdimen2\font plus
\BIBentryALTinterwordstretchfactor\fontdimen3\font minus
  \fontdimen4\font\relax}
\providecommand{\BIBforeignlanguage}[2]{{%
\expandafter\ifx\csname l@#1\endcsname\relax
\typeout{** WARNING: IEEEtran.bst: No hyphenation pattern has been}%
\typeout{** loaded for the language `#1'. Using the pattern for}%
\typeout{** the default language instead.}%
\else
\language=\csname l@#1\endcsname
\fi
#2}}
\providecommand{\BIBdecl}{\relax}
\BIBdecl

\bibitem{FazeliVardyYaak}
A.~Fazeli, A.~Vardy, and E.~Yaakobi, ``{PIR} with low storage overhead: Coding
  instead of replication,'' \emph{CoRR}, vol. abs/1505.06241, 2015.

\bibitem{VardyYaak}
A.~Vardy and E.~Yaakobi, ``Constructions of batch codes with near-optimal
  redundancy,'' in \emph{{IEEE} International Symposium on Information Theory,
  {ISIT}}, 2016, pp. 1197--1201.

\bibitem{ChorGoldreichKushilevitzSudan_ACM}
B.~Chor, E.~Kushilevitz, O.~Goldreich, and M.~Sudan, ``Private information
  retrieval,'' \emph{J. {ACM}}, vol.~45, no.~6, pp. 965--981, 1998.

\bibitem{ShahRR14}
N.~B. Shah, K.~V. Rashmi, and K.~Ramchandran, ``One extra bit of download
  ensures perfectly private information retrieval,'' in \emph{{IEEE}
  International Symposium on Information Theory {ISIT}}, 2014, pp. 856--860.

\bibitem{AugotLS14}
D.~Augot, F.~Levy{-}dit{-}Vehel, and A.~Shikfa, ``A storage-efficient and
  robust private information retrieval scheme allowing few servers,'' in
  \emph{Cryptology and Network Security - 13th International Conference,
  {CANS}}, 2014, pp. 222--239.

\bibitem{ChanHY15}
T.~H. Chan, S.~Ho, and H.~Yamamoto, ``Private information retrieval for coded
  storage,'' in \emph{{IEEE} International Symposium on Information Theory,
  {ISIT} 2015, Hong Kong, China, June 14-19, 2015}, 2015, pp. 2842--2846.

\bibitem{FazeliVardyYaakISIT}
A.~Fazeli, A.~Vardy, and E.~Yaakobi, ``Codes for distributed {PIR} with low
  storage overhead,'' in \emph{{IEEE} International Symposium on Information
  Theory, {ISIT}}, 2015, pp. 2852--2856.

\bibitem{RaoVardy}
S.~Rao and A.~Vardy, ``Lower bound on the redundancy of {PIR} codes,''
  \emph{CoRR}, vol. abs/1605.01869, 2016.

\bibitem{BlackburnEtzion}
S.~R. Blackburn and T.~Etzion, ``{PIR} array codes with optimal {PIR} rate,''
  \emph{CoRR}, vol. abs/1607.00235, 2016.

\bibitem{ZhangWangWeiGe}
Y.~Zhang, X.~Wang, H.~Wei, and G.~Ge, ``On private information retrieval array
  codes,'' \emph{CoRR}, vol. abs/1609.09167, 2016.

\bibitem{Muller}
D.~E. Muller, ``Application of boolean algebra to switching circuit design and
  to error detection,'' \emph{Trans. {I.R.E.} Prof. Group on Electronic
  Computers}, vol.~3, no.~3, pp. 6--12, 1954.

\bibitem{Reed}
I.~S. Reed, ``A class of multiple-error-correcting codes and the decoding
  scheme,'' \emph{Trans. of the {IRE} Professional Group on Information Theory
  {(TIT)}}, vol.~4, pp. 38--49, 1954.

\bibitem{Lachaud86}
G.~Lachaud, ``Projective reed - muller codes,'' in \emph{Coding Theory and
  Applications, 2nd International Colloquium}, 1986, pp. 125--129.

\end{thebibliography}

\end{document}